\documentclass{moriond}

\def\be{\begin{equation}}
\def\ee{\end{equation}}
\def\bea{\begin{eqnarray}}
\def\eea{\end{eqnarray}}

\newcommand{\Photo}{}

\begin{document}
\vspace*{1cm}
\title{Searches for long-duration gravitational wave
transients in LIGO and Virgo data}

\author{A. Macquet$^1$, M.A. Bizouard$^1$, N. Christensen$^1$ and M. Coughlin$^2$}
\address{ $^1$Artemis, Universit\'e C\^ote d’Azur, Observatoire de la C\^ote d’Azur, CNRS, Nice 06300, France}
\address{$^2$School of Physics and Astronomy, University of Minnesota, Minneapolis, Minnesota 55455, USA}

\maketitle\abstracts{
Long-lived gravitational wave (GW) transients have received interest in the last decade, as the sensitivitity of LIGO and Virgo. Such signals, lasting between $10^1 - 10^3$s, can come from a variety of sources, including accretion disk instabilities around black holes, binary neutron stars post-merger, core-collapse supernovae, non-axisymmetric deformations in isolated neutron stars, and magnetar giant flares. Given the large parameter space and the lack of precisely modeled waveforms, searches must rely on robust detection algorithms, which make few or no assumption on the nature of the signal. 
Here we present a new data analysis pipeline to search for long-lived transient gravitational wave signals, based on an excess cross-power statistic computed over a network of detectors. It uses a hierarchical strategy that allows to estimate the background quickly and implements several features aimed to increase detection sensitivity by $\sim 30 \%$ for a wide range of signal morphology compared to an older implementation. We also report upper limits on the GW energy emitted from a search conducted with the pipeline for GW emission around a sample of nearby magnetar giant flares, and discuss detection potential of such sources with second and third generation detectors.
}

\section{Presentation of an unmodeled GW search pipeline}

When the GW signal searched is precisely modelled, e.g for compact binary mergers or cosmic strings, searches usually rely on matched filter techniques, which have optimal sensitivity. However, detection of GW transients which are poorly modelled, often refered as GW "bursts", require a more agnostic approach. Most of the searches rely on finding zones of excess power in time-frequency representation of the data. This is the case of the \texttt{STAMP} pipeline\cite{STAMP}, for which we propose here an enhanced version, \texttt{PySTAMPAS}, designed to perform all-sky / all-time GW searches at a reduced computational cost and increased detection efficiency\cite{Macquet:2021}.

\subsection{Coherent detection statistic}
We consider a pair $(I,J)$ of GW detectors, and their strain time-series $s_I(t)$ and $s_J(t)$. A simple time-frequency representation of the data can be obtained by taking the Fourier transform of short segments of duration $\Delta T$. We note $\tilde{s}_I(t;f)$ such a representation, where $t;$ denotes the start time of each segment.
Following \cite{STAMP}, an estimator for the coherent GW power signal-to-noise ratio (SNR) in a time-frequency pixel $(t,f)$ is
\begin{equation}
\label{eq:SNR}
  \rm{SNR}(t;f,\hat{\Omega})
  = Re \left[\frac{2 \, \tilde{s}_{I}^{\star}(t;f)\tilde{s}_{J}(t;f)} {\sqrt{\frac{1}{2} P_I(t;f)P_J(t;f)}} ~ e^{2 \pi i f \hat{\Omega} \cdot \Delta \vec{x}_{IJ}/c} \right],
  \end{equation}
where $P_I(t;f)$ is the power spectral density of the detector's noise and $\Delta \vec{x}_{IJ}$ the distance between the two detectors. The GW signal is assumed to originate from a point-like source whose direction is given by the unit vector $\hat{\Omega}$. The phase term comes from the delay between the two detectors of the arrival of the GW signal.

When doing an all-sky search, the direction of the source is not known \textit{a priori}, introducing a degeneracy in the computation of $\textrm{SNR}(t;f, \hat{\Omega})$. However, a long-lived GW can spread over several pixels, forming a cluster $\Gamma$. This breaks the degeneracy as $\hat{\Omega}$ remains identical for each pixel. By testing a sufficient number of sky positions, we maximize the summed SNR of the cluster
\begin{equation}
\textrm{SNR}_{\Gamma}(\hat{\Omega}) \equiv \sum \limits_{(t;f) \in \Gamma} \textrm{SNR}(t;f,\hat{\Omega}).
\end{equation}
Noise fluctuations in detectors can form clusters that mimic the shape of a GW signal. To better discriminate signal from noise, we assign each cluster a ranking statistic $p_{\Lambda}$ that reflects its significance, based on the coherent and incoherent components of the SNR of the pixels\cite{Macquet:2021}.

\subsection{Implementation of the detection algorithm}

Because of the large parameter space and the amount of data to process, unmodeled searches are computationaly expensive, and  compromises have to be made between detection efficiency and computational cost, leading to suboptimal sensitivity. \texttt{PySTAMPAS} implements a hierarchical method \cite{Thrane:2015psa} to address computational cost issues while preserving detection sensitivity. The different steps of these methods are described below.

\paragraph{Time-frequency maps}
The data stream from each detector is split into windows of $\sim 500$s duration, and for each window, time-frequency maps ($tf$-maps) of the detectors' strain data are built using short segments of duration $\Delta T$. The segments are hann-windowed and overlap by $50\%$. In order to better reconstruct fast frequency evoluting signals, different values of $\Delta T$ are used and the resulting maps are combined to form a multi-resolution $tf-map$.
The maps are whitened by the one-sided amplitude spectral density $\sqrt{P_I(t;f)}$, which is estimated by
taking the median of $|\tilde{s}_I(t;f)|^2$ over frequency neighbouring pixels.

\begin{figure}
\label{ftmap}
\includegraphics[scale=0.28]{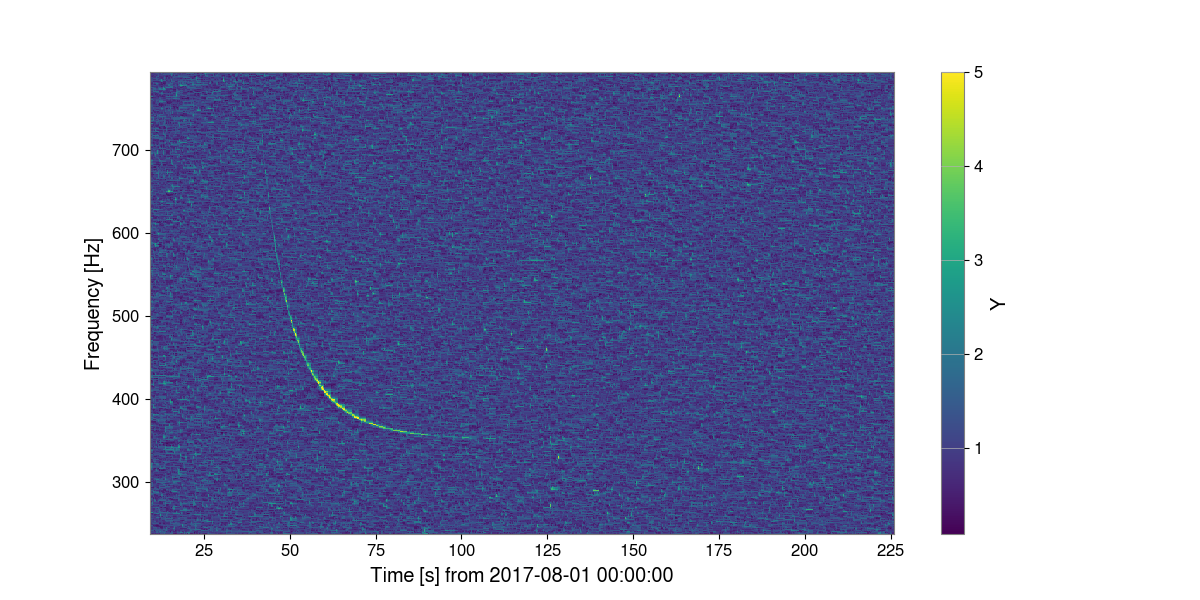}
\includegraphics[scale=0.28]{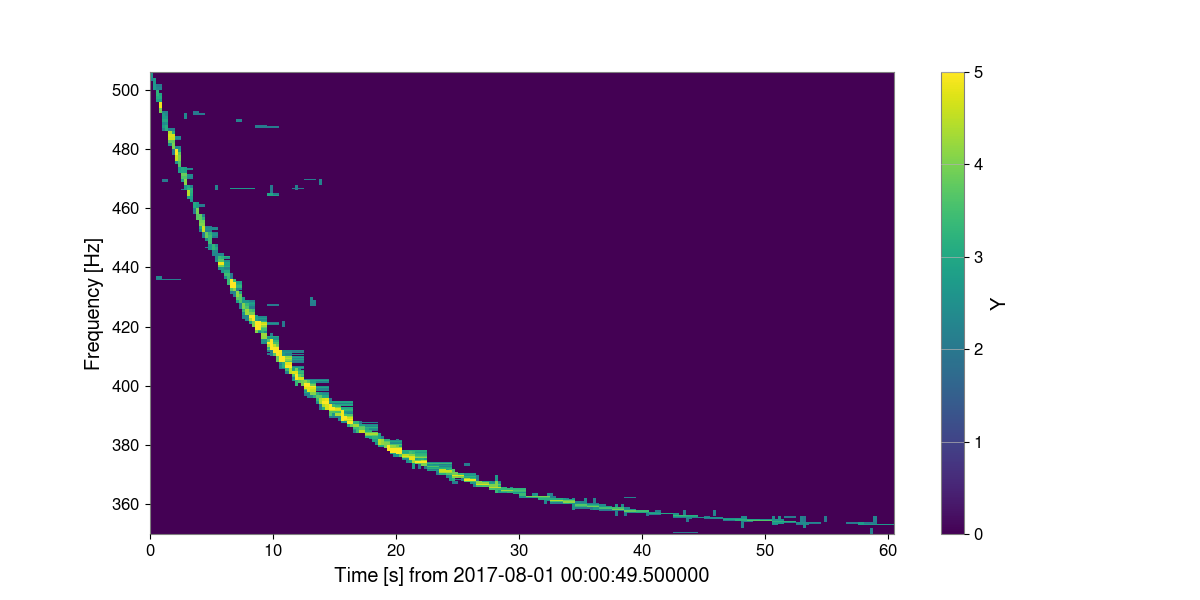}
\caption{\textit{Left} : multi-resolution time-frequency map of a long-duration GW signal injected in Gaussian data. \textit{Right} : cluster extracted from this $tf$-map by the clustering algorithm.}
\end{figure}

\paragraph{Clustering}
A statistically significant GW signal will generate a track of excess power pixels in a $tf$- map. To identify and extract those tracks, a clustering algorithm is run over the maps. Several types of clustering algorithms have been implemented for un-modelled GW searches, which are adapted for different classes of signal. To remain as agnostic as possible on the signal morphology, we opt for a seed-based clustering algorithm. The \texttt{burstegard} algorithm\cite{Prestegard:2016} identifies pixels of single detector $tf$-maps with autopower above a given threshold and groups the neighbouring pixels together to form clusters (see Fig. \ref{ftmap}). Cluster which contain a sufficient number of pixels are kept to be analyzed coherently at the next step of the analysis. 
\paragraph{Coincident analysis}
Clusters extracted from one detector are matched with corresponding pixels from the other detector's $tf$-map. Assuming a sky direction $\hat{\Omega}$, the coherent SNR of each pixel is computed following eq.\ref{eq:SNR}. Because cross-correlation is only computed over a reduced number of pixels, it is possible to test many sky positions (up to several hundreds), and therefore maximize the summed quantity $\textrm{SNR}_{\Gamma}$ while keeping computational cost affordable. The coincident triggers extracted are ranked by significance following their coherent statistic $p_{\Lambda}$.

\paragraph{Background estimation}

To assess a false-alarm probability to triggers found in coincidence, one has to estimate the distribution of $p_{\Lambda}$ for triggers generated by detectors' noise (background noise). Like a lot of unmodeled searches, \texttt{PySTAMPAS} uses the time-slide method to estimate background noise distribution. The principle is to shift the data stream from one detector with respect to the other in order to remove any coherent GW signal from the data, while preserving any non Gaussian or non-stationary feature that can be present in real GW detector's data. By repeating the process for multiple values of time-shift, it is possible to simulate a large amount of background noise and therefore assess precisely the significance of coincident triggers. With the hierarchical method implemented, \texttt{PySTAMPAS} is able to estimate the significance of triggers up to $5\, \sigma$ for a year-long observing run. 

\subsection{Performances}
We estimate the performances of the pipeline by injecting simulated signals in the data at different amplitudes and try to recover them. To simulate a realistic all-sky search, detectors' response to each signal is computed assuming a random sky localization and source's plane inclination. We considered a set of a dozen waveforms that span the parameter space, with duration and frequency  between $8-240$ s and $10-1800$ Hz respectively, and diverse spectral morphology. Results on Gaussian simulated data and real data from Advanced LIGO's second observing run show a detection sensitivity increased by $\sim 30 \%$ in average for the set of waveforms tested, as compared to the previous all-sky implementation of the \texttt{STAMP} pipeline.

\section{Search for long-duration GW emission around Magnetar Giant Flares}
\subsection{Motivation}
Magnetar Giant Flares (MGF) are highly energetic events that occur in magnetars, young neutron stars with very large magnetic field (up to $10^{14}$ G). They usually consist in a bright millisecond flare of gamma-rays of energy $10^{44-47}$ erg followed by pulsating tails at $\sim 10^{44}$ erg that can last several hundred of seconds. 

The recent observation of GRB 200415a, a short gamma-ray burst (GRB) in NGC 253 and its possible magnetar giant flare origin\cite{2021ApJ...907L..28B}, along with $3$ other short GRBs from nearby galaxies, has motivated us to perform a search for a GW counterparts around those events with \texttt{PySTAMPAS}. Indeed, the large energy emission at relatively close distance ($< 5$ Mpc) could make them observable with GW, and the lack of knowledge about the coupling between oscillation modes and GW emission require an un-modelled search algorithm.

\subsection{Data sample and search methodology}
We considered $3$ short GRBs associated with MGF that happened between 2005 and 2007 for which data from initial LIGO detectors at Hanford (H1, H2) and Livingston (L1) were available\footnote{LIGO and Virgo detectors were off at the time of GRB 200415a}. Since a host galaxy has been identified for each of these events, \texttt{PySTAMPAS} pipeline was used in a targeted configuration, with only one sky position tested per search. Searches were conducted over a $[-500\,\rm{s}, +500\,\rm{s}]$ around each event to account for the potential delay between EM and possibly long-duration GW emission.
\subsection{Results and future prospects}

No significant GW candidates were found for any of the three events searched. To place upper limits on the GW energy emitted we injected sinusoidal waveforms damped with an exponential enveloppe with decay times between $0.2-10$ s and central frequency between $50-500$ Hz. For two of the events, GRB 051103 and GRB 070222, we derive upper limits at $100$ Hz of $6 \times 10^{51}$ erg and $3 \times 10^{50}$ erg respectively. These are within the same order of magnitude as upper limits previously published in \cite{abadie2012implications}, with a factor 2 improvement. We also place an upper limit of $10^{52}$ erg for GRB 070222, which had only been recently associated with a potential MGF. Detailed results can be found in \cite{MGF}.

\begin{figure}
\centering
\label{MGFplot}
\includegraphics[scale=0.27]{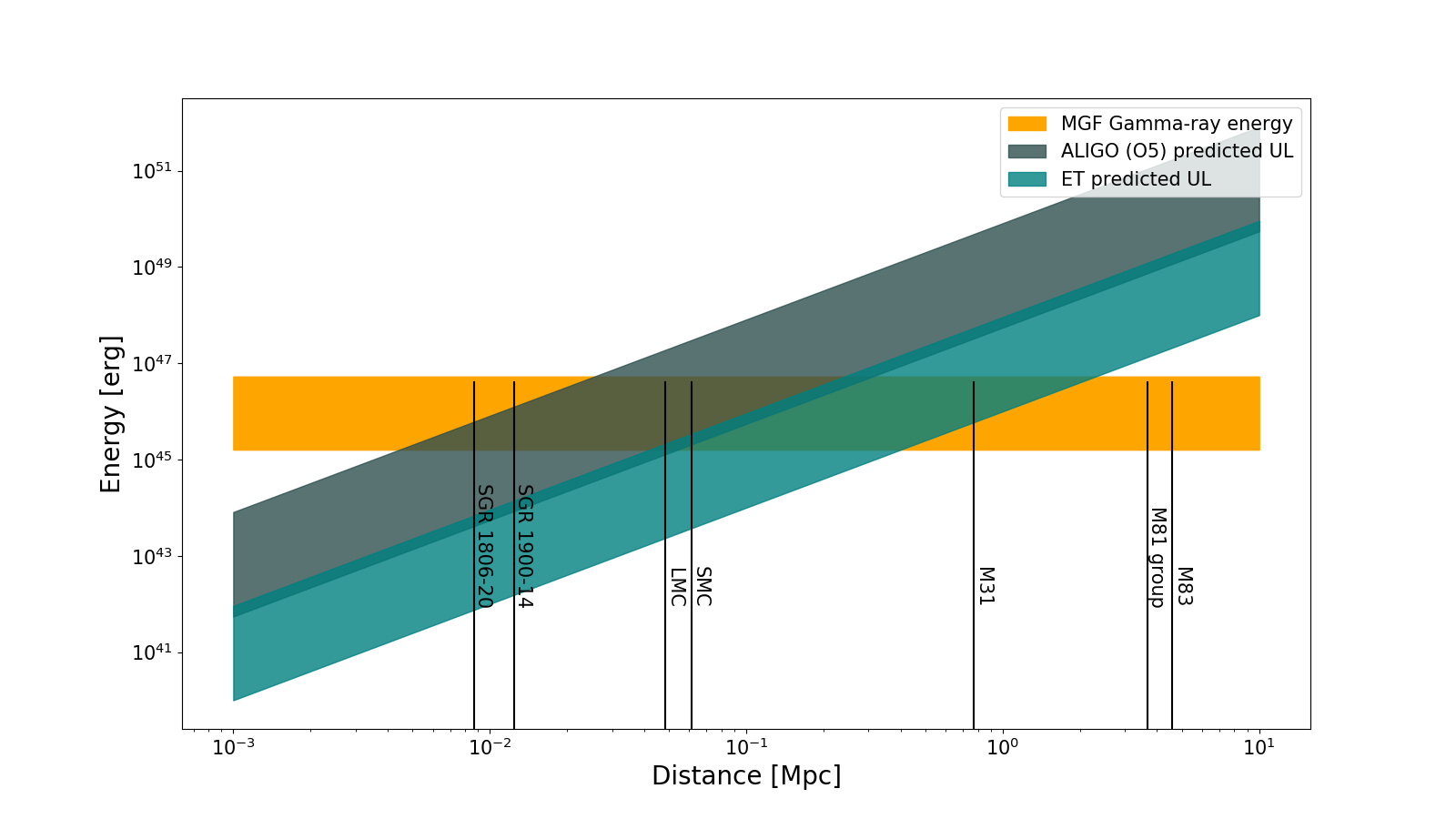}
    \caption{Estimated sensitivity of \texttt{PySTAMPAS} to GW energy emitted vs distance to the source for Advanced LIGO at design sensitivity (grey band) and Einstein Telescope (yellow band). The yellow band shows the typical electromagnetic energy emitted by MGFs for comparison.}

\end{figure}

We also studied the prospectives of future detections using simulated data of Advanced LIGO at design sensitivity (expected during O5) and of Einstein Telescope. We report the results in Fig. \ref{MGFplot} that shows in particular that galactic MGF should be observable with Advanced LIGO if they emit up to $1 \%$ of their EM energy in GW. Furthermore,  Third generation detectors could push these boundaries further and make even less energetic MGFs (potentially associated with SGR and FRB) within detection range.

\section*{References}

\bibliography{bibliography}
\bibliographystyle{unsrt}  

%
%
%
%
%

\end{document}